\documentclass[notoc,paper,letterpaper]{JHEP3}
\usepackage{graphicx}
\usepackage{amssymb,amsmath,cite}

\newcommand{\be}{\begin{equation}} 
\newcommand{\ee}{\end{equation}} 
\newcommand{\bea}{\begin{eqnarray}} 
\newcommand{\eea}{\end{eqnarray}} 
\def\Z{\mathbb{Z}} 

\def\til{\widetilde}
\def\bar{\overline}
\def\a{{\alpha}}
\def\b{{\beta}}
\def\g{{\gamma}}
\def\d{{\delta}}
\def\L{{\Lambda}}
\def\ff{{\mathfrak f}}

\def\gf{{\mathfrak g}}
\def\tgf{{\til\gf}}
\def\hf{{\mathfrak h}}

\def\ef{\mathfrak{e}}
\def\uf{\mathop\mathfrak{u}} 
\def\suf{\mathop\mathfrak{su}} 
 
\def\sof{\mathop\mathfrak{so}} 
 
\def\spf{\mathop\mathfrak{sp}} 

\title{Mass deformations of four-dimensional, rank 1, N=2 superconformal field theories}
\author{Philip C. Argyres\\
Physics Department, University of Cincinnati, Cincinnati OH 45221-0011, USA\\
E-mail: \email{philip.argyres@gmail.com}}
\author{John Wittig\\
John's address\\
E-mail: \email{jwittig0126@gmail.com}}
\abstract{
Turning on N=2 supersymmetry-preserving relevant operators in a 4-dimensional N=2 superconformal field theory (SCFT) corresponds to a complex deformation compatible with the rigid special K\"ahler geometry encoded in the low energy effective action.  Field theoretic consistency arguments indicate that there should be many distinct such relevant deformations of each SCFT fixed point.  Some new supersymmetry-preserving complex deformations are constructed of isolated rank 1 SCFTs.  We also make predictions for the dimensions of certain Higgs branches for some rank 1 SCFTs.}

\begin{document}
\section{Introduction}
In this paper we discuss the existence and construction of N=2 supersymmetric mass deformations of rank 1 SCFTs.  Here  by ``rank" we mean the complex dimension of the Coulomb branch of the SCFT.  It corresponds to the rank of the gauge group for Lagrangian theories, and generalizes that concept to the non-Lagrangian theories that will be our focus.  After reviewing the classification of rank 1 N=2 SCFTs in section 2, we will see that even at rank 1 the classification and construction of all possible N=2 mass deformations is a difficult and open algebraic problem.  In section 3 we describe a direct approach to the problem and illustrate its difficulties, then turn in section 4 to a less systematic but more fruitful approach.  We will be able to construct a new N=2 mass deformation of the $E_6$ conformal curve which has a $G_2$ global flavor symmetry.  In sections 5 and 6 we will compare this result to predictions following from strong coupling dualities of Lagrangian N=2 SCFTs.  In particular, in section 6, we will refine some earlier predictions by including information about the dimension of certain Higgs branches at N=2 fixed points.

\section{Known mass deformations of rank 1 N=2 SCFTs} 

Although a classification of rank 1 N=2 SCFTs is known \cite{apsw9511,mn9608}, a complete list of the possible mass deformations of those theories is not known.  We start by reviewing the classification of the rank 1 N=2 SCFTs.  This is done through the construction of their Seiberg-Witten curves \cite{sw9408}.  These scale-invariant curves are the starting point for turning on mass deformations which break the scale invariance.
A general picture that has emerged \cite{as0711,aw0712} and is a useful orientation for discussing the curves of strongly-coupled (non-Lagrangian) N=2 supersymmetric fixed-point theories is that strongly coupled conformal fixed points have a similar structure as Lagrangian theories:
\begin{itemize}
\item the singularity of the Seiberg-Witten curve plays a role analogous to the gauge algebra ($\gf$) of the Lagrangian theory, and
\item different complex structure deformations of the singular curve that preserve N=2 supersymmetry are analogous to adding matter hypermultiplets in different representations ($\bf r$) of $\gf$ that preserve conformal invariance (in the limit of zero masses).
\end{itemize}


The low-energy physics on the Coulomb branch can be encoded by a family of elliptic curves \cite{sw9408},
\be
y^2 = x^3 + f(u,m_i)\,x + g(u,m_i)
\ee
depending on complex parameters $\{ u, m_i\}$,
and a meromorphic 1-form $\lambda_{SW}$ with residues $m_i$ at its poles and which satisfies 
\be\label{e2}
\partial_u\lambda_{SW} = y^{-1}{\rm d}x + \partial_x(\star){\rm d}x.
\ee
Here $u$ is the a global complex coordinate on the Coulomb branch with scaling dimension $D(u)$, and $m_i$, $i=1,\ldots,h$, are mass parameters, all of scaling dimension $D(m_i)=1$.  

When the $m_i=0$, the theory is scale invariant.  In particular, the scale-invariant vacuum corresponds to the origin, $u=0$, of the Coulomb branch.  In vacua with $u\neq0$ the scale invariance is spontaneously broken.  The scale-invariant theory has  a global internal symmetry algebra $\uf(2)_R\oplus\hf$.  Here $\uf(2)_R$ is the R-symmetry and $\hf$ is the global flavor symmetry.  Turning on masses $m_i$ explicitly breaks the flavor symmetry.  The masses transform as weights in the adjoint representation of $\hf$, so turning on generic $m_i$ breaks $\hf \to \oplus_{i=1}^h \uf(1)_i$, where $h={\rm rank}(\hf)$.  Charges, $n_i$, of states under these $\uf(1)$'s are called quark numbers.

The complex structure of the torus described by the curve is the low energy $\uf(1)_{em}$ gauge coupling.  The electric and magnetic $\uf(1)_{em}$ charges $(n_e,n_m)$ and quark numbers $n_i$ of a BPS state determine the homology class of a cycle, $\gamma(n_e,n_m,n_i) = n_{e}[{\alpha}] + n_{m} [{\beta}]+n_i[{\delta_i}]$, on the torus, which determines the central charge (and BPS mass) of these states by ${\rm Z} \ = \ \oint_{\gamma}\lambda_{SW}$.  Here $\alpha$ and $\beta$ are a canonical basis of 1-cycles on the torus, and the $\delta_i$ are a basis of cycles around the poles of of $\lambda_{SW}$.

The elliptic curve is singular at values of $u$ corresponding to the zeros of the discriminant 
\be
\Delta \equiv 4\cdot f^3 + 27\cdot g^2 = 0.
\ee
These singularities physically correspond to points on the Coulomb branch where $\uf(1)_{em}$-charged states are becoming massless.  For a curve representing a scale-invariant theory (with masses $m_i=0$), then $\Delta \sim  u^{n}$ for some $n$.  When mass parameters are turned on they appear in the curve in the form of Weyl invariants (adjoint Casimirs), $M_a$, of the flavor symmetry, where the $M_a$ are homogeneous polynomials of degree $a$ in the $m_i$ and the $a$ are the exponents plus one of the flavor Lie algebra $\hf$.  Then the discriminant will take the form
\be
\Delta = u^{n}+...+P_k(\{M_a\})u^{n-k}+\cdots ,
\ee
where the $P_k$ are homogeneous polynomials in the $m_i$ of appropriate degree.
Since the order of $\Delta$ is independent of the flavor symmetry, this implies that different flavor symmetries correspond to different patterns of orders of zeros of $\Delta$ in $u$.


The possible scale-invariant singularities of rank-1 curves coincides with Kodaira's classification \cite{kodaira} of the degenerations of holomorphic families of elliptic curves over one variable \cite{mn9608}.  The result is two infinite series and six ``exceptional" curves.
\be\begin{array}{lllcc}
\mbox{singularity\ \ \ } & \mbox{curve} & \Delta & D(u) & \gf \\ \hline
E_8 & y^2 = x^3 + 2 u^5    & u^{10}      & 6   & -  \\
E_7 & y^2 = x^3 + u^3 x    & u^9         & 4   &  -  \\
E_6 & y^2 = x^3 + u^4      & u^8         & 3   &  -  \\
D_4 & y^2 = x^3 + 3 \tau u^2 x + 2 u^3 & u^6 
& 2 & \suf(2) \\
H_3 & y^2 = x^3 + u^2      & u^4         & 3/2 &  -  \\
H_2 & y^2 = x^3 + u x      & u^3         & 4/3 & - \\
H_1 & y^2 = x^3 + u        & u^2         & 6/5 & - \\
D_{n>4} & y^2 = x^3 + 3 u x^2 + 4 \Lambda^{-2(n-4)} u^{n-1}\ \    
& u^{n+2}\ \ 
 & \ \ \ 2\ \ \  &  \suf(2)  \\
A_{n\ge0} & y^2 = (x-1) (x^2 + \Lambda^{-(n+1)} u^{n+1} ) 
& u^{n+1} 
& 1   &  \uf(1) 
\end{array}\label{e5}
\ee
The ``singularity" column gives the conventional names for each singular family of curves; the Lie algebra-like A-D-E nomenclature comes from a correspondence between the A-D-E affine Lie algebra extended Dynkin diagrams and the pattern of blowups resolving those singularities.  $\gf$ is the gauge algebra for the curves in the list corresponding to Lagrangian theories,
$\Lambda$ is the UV strong coupling scale of the IR-free CFTs, and $\tau$ is the marginal gauge coupling.  The other six curves correspond to strongly interacting fixed point theories.
The $H_{1,2,3}$ series were found by tuning parameters in $\suf(2)$ and $\suf(3)$ Lagrangian field theories in \cite{ad9505,apsw9511}, while the $E_{6,7,8}$ curves were constructed in \cite{mn9608}. 

The maximal mass deformations of these curves correspond 
to the most general complex structure deformations which do not increase the order of the singularity of the discriminant for any non-zero values of the deformation parameters:
\be\begin{array}{ll}
\mbox{sing.} & \mbox{maximally deformed curve:\ \ \ } y^2\ =\ \ldots \hfill \hf \\ \hline
E_8 & x^3 + x(M_2u^3+M_8u^2+M_{14}u+M_{20})
+ (2u^5+M_{12}u^3+M_{18}u^2+M_{24}u+M_{30}) \hfill\ \ \ \ef_8 \\
E_7 & x^3 + x(u^3+M_8u+M_{12}) + (M_2u^4+M_6u^3+M_{10}u^2
+M_{14}u+M_{18}) \hfill \ef_7 \\
E_6 & x^3 + x(M_2u^2+M_5u+M_8) + (u^4+M_6u^2+M_9u+M_{12}) \hfill \ef_6 \\
D_4 & x^3 + x(3 \tau u^2+M_2u+M_4) + (2u^3+\til M_4u+M_6) \hfill \sof(8) \\
H_3 & x^3 + x(M_{1/2}u+M_2) + (u^2+M_3) \hfill \uf(3) \\
H_2 & x^3 + x(u) + (M_{2/3}u+M_2) \hfill \uf(2) \\
H_1 & x^3 + x(M_{4/5}) + (u) \hfill \uf(1) \\
D_{n>4} & x^3 + 3 u x^2 + \L^{-(n-4)} \til M_n x
+ 4 \L^{-2(n-4)} (u^{n-1}+M_2u^{n-2}+\cdots+M_{2n-2}) \hfill \sof(2n) \\
A_{n\ge0}\ \ & (x-1) (x^2 + \L^{-(n+1)} [u^{n+1} + M_2 u^{n-1}+
M_3 u^{n-2} + \cdots + M_{n+1}]) \hfill \suf(n+1)\\
\end{array}\label{e6}
\ee
Here the Lie algebra on the right is the flavor algebra, $\hf$, of the corresponding SCFT.  These deformations are all compatible with the requirement of N=2 supersymmetry; {\it e.g.}, one can show that there exists a meromorphic 1-form, $\lambda_{SW}$, satisfying (\ref{e2}).  

Note that for the A-D-E series of curves, the maximal flavor algebras are the same as the Lie algebras used to name the singularities.  But, as we will now discuss, this is just a coincidence.

As an example, consider the maximal mass deformation of the $A_n$ series which corresponds to a $\uf(1)$ gauge theory with $n+1$ hypermultiplets of charge 1, contributing $b=n+1$ to the beta function.  The total contribution to the beta function from the hypermultiplets determines the form of the singularity in an IR-free theory, but there are many inequivalent ways of contributing a given amount.  For example, $n_{a}$ hypermultiplets of charge $\pm r_{a}$ give $b = \sum_{a}n_{a}r_{a}^{2}$, and gives a $\oplus_{a}\uf(n_{a})$ flavor symmetry, smaller than the $\suf(n+1)$ flavor symmetry of the maximal mass deformation.  Thus there should be many inequivalent mass deformations of the $A_n$ singularities preserving N=2 supersymmetry.

The same is true for the other Lagrangian curves,  the $D_n$ series which are IR-free $\suf(2)$ theories for $n>4$, and the conformal $\suf(2)$ theory for $n=4$.
For the $D_4$ theory we can have $b=0$ with 4 fundamental hypermultiplets or with 1 adjoint hypermultiplet.  The first is the maximal mass deformation and has an $\sof(8)$ flavor symmetry, while the second has an $\spf(1)$ flavor symmetry, with curve with a sub-maximal mass deformation \cite{sw9408}
\be
y^{2}=\prod_{i}(x-e_{i}u-e_{i}^{2}M_{2}).
\ee

The mass deformation of the $D_{n>4}$ curves in (\ref{e6}) corresponds to an $\suf(2)$ gauge theory with $2n$ half-hypermultiplets in the fundamental representation contributing $b = 2(n-4)$.
(The total $b > 0$, so all these theories are IR free.)  There are many other ways of adding matter hypermultiplets to contribute the same $b$.  There are two classes of representations for $\suf(2)$, the real $\bf{2r+1}$ and the pseudoreal $\bf{2s}$ (denoting representations by their dimensions).  To avoid anomalies \cite{w82} we must have $2n_{r}$ of each real representation and any number $m_{s}$ of the pseudoreal such that their contribution to the appropriately normalized quadratic casimir, $\frac{1}{3} \sum_s m_s s(4s^2-1)$, is even.  Then $b = \frac{4}{3} \sum_r n_r r(r+1)(2r+1)+\frac{1}{3} \sum_s m_s s(4s^2-1)-8$
and the flavor symmetry that corresponds to this value of the beta function is
$\oplus_{r}\spf(n_r) \oplus_{s}\sof(m_{s})$.  Again, we see that there are many inequivalent sub-maximal mass deformations of the $D_{n>4}$ singularities that preserve N=2 supersymmetry.

\section{Constructing new curves}

So it is natural to try to construct sub-maximal mass deformations of the non-Lagragian $E_{6,7,8}$ and $H_{1,2,3}$ singularities.   Note that evidence for the existence of new mass deformations of the $E_{6,7,8}$ singularities was found in \cite{aw0712} and will be reviewed in section 5.  The maximal mass deformations of the $H_{1,2,3}$ singularities were found in \cite{ad9505,apsw9511} by taking appropriate scaling limits of known Lagrangian curves, while the maximal mass deformations of the $E_{6,7,8}$ curves were worked out by Minahan and Nemeschansky in \cite{mn9608}
just from guessing the flavor symmetry and imposing the conditions of N=2 supersymmetry.  We will try to copy that procedure below.   

To illustrate the problem, suppose we look for a sub-maximal mass deformation of the $E_6$ singularity with flavor symmetry $\hf=\suf(3)$.  Let us furthermore suppose that the discriminant, $\Delta$, will have only $Z=4$ zeros in the $u$-plane for generic $m_i$  instead of the maximal 8.   We have no reason for supposing that such a mass deformation should exist; a systematic search for all sub-maximal deformations would require a similar calculation for the whole list of relevant $\hf$ and $Z$.  (In sections 5 and 6 below we will review the flavor symmetries, $\hf$, of sub-maximal mass deformations predicted to exist by S-duality arguments, and will extend the arguments of \cite{st0804} to determine the number of generic zeros, $Z$, of $\Delta$ associated with these $\hf$.)  $Z=4$ can only be achieved by one of the following possible inequivalent factorizations of the discriminant of the $E_6$ singular curve:
\bea
\Delta &\sim& (u+...)^5 (u^3+...)\quad \mbox{or}\quad 
(u+...)^4 (u+...)^2 (u^2+...)\quad \mbox{or}\quad
(u+...)^3 (u+...) (u^2+...)^2\nonumber\\
&&\quad \mbox{or} \quad
(u^2+...)^3 (u^2+...)\quad \mbox{or}\quad
(u^4+...)^2.
\eea
A laborious computer search of parameterizations of complex deformations of the $E_6$ singularity reveals 2 consistent factorization solutions.
The first $\suf(3)$ factorization solution is actually a 1-parameter ({$\nu$}) family:
\bea
y^{2} &=& x^{3} + 3 N_{2} x[u^{2}+(1+{\nu})N_{2}^{3}+N_{3}^{2}] + 
[u^{4}+u^{2}((1+2 {\nu})N_{2}^{3}+2 N_{3}^{2})\nonumber\\
&& \qquad \mbox{}+
{\nu}(1+ {\nu})N_{2}^{6}+(1+2 
{\nu})N_{2}^{3}N_{3}^{2}+N_{3}^{4}],\\
\Delta &=& -27[u^{2}+(1+{\nu})N_{2}^{3}+N_{3}^{2}]^{2}[u^{2}+(2+{\nu})N_{2}^{3}+N_{3}^{2}]^{2} \ \sim\  (u^4+\ldots)^2,\nonumber
\eea
where $N_{2,3}$ are the $\suf(3)$ Weyl invariants.
However, there does not exist a Seiberg-Witten 1-form for this curve for any {$\nu$}, so these deformations are not compatible with $N=2$ supersymmetry.
The second $\suf(3)$ factorization solution is
\bea
y^{2} &=& x^{3} + u[3N_{2}x(u-4N_{3})+u^{3}-12u^{2}N_{3}-u(N_{2}^{3}-48N_{3}^{2})-64N_{3}^3] \nonumber\\
\Delta &=& -27u^{2}[u^{3}-12u^{2}N_{3}+u(N_{2}^{3}+48N_{3}^{2})-64N_{3}^{3}]^{2}
\nonumber
\eea
We have not been able to construct or rule out a Seiberg-Witten 1-form for this curve.

As another example, we can search for a sub-maximal deformation of the $E_7$ singularity with flavor symmetry $\spf(3) \oplus \suf(2)$ (as predicted to exist in \cite{aw0712}).  This requires factoring a 9th-order polynomial into $Z=6$ generic zeros (as predicted in section 6 below):
\be\label{e10}
\Delta \sim (u+...)^{4}(u^{5}+...)\quad\mbox{or}\quad
(u+...)^{3}(u+...)^{2}(u^{4}+...)\quad\mbox{or}\quad
(u^{3}+...)^{2}(u^{3}+...).
\ee
A systematic search reduces the problem to solving on the order 
of $800$ polynomial relationships among $160$ unknowns.   Though highly over-constrained, it is computationally very difficult to determine whether there are any solutions.

Similarly, the sub-maximal deformation of the $E_8$ singularity with flavor symmetry
$\spf(5)$ predicted by \cite{aw0712} requires factoring a 10th-order polynomial into $Z=7$ generic zeros as predicted in section 6 below:
\be\label{e11}
\Delta \sim (u+...)^{4}(u^{6}+...) \quad\mbox{or}\quad
(u+...)^{3}(u+...)^{2}(u^{5}+...)\quad\mbox{or}\quad
(u^{3}+...)^{2}(u^{4}+...).
\ee
A systematic search in this case is computationally much too difficult.  Clearly some other method is needed.

\section{Isogenies: a non-systematic approach}

An $n$-isogeny is an $n$-to-1 holomorphic map of a curve to itself that preserves the holomorphic 1-form.  The existence of such a map implies that some of the zeros of the discriminant will have multiplicity $n$.  Though there are, in principle, $n$-isogenies for any $n$, only certain $2$- and $3$-isogenies seem to have known closed-form expressions for general parameters \cite{h04}.   In particular, there are three
traditional presentations of elliptic curves which are related by simple isogenies:
\bea
\mbox{Legendre:} &\quad& y^{2} = x^{3} + f x + g, \nonumber\\
\mbox{Jacobi:} && \til y^{2} = \til x^{4} + \alpha \til x^{2} + \beta, \\
\mbox{Hessian:} && \gamma = \til y^{3} + \delta \til x\til y + \til x^{3}, \nonumber
\eea
where $f$, $g$, $\alpha$, $\beta$, $\gamma$, and $\delta$ are all functions of $u$.
We have been presenting our elliptic curves in the Legendre form so far.  The map between the Jacobi and Legendre forms is a 2-isogeny, while that between the Hessian and Legendre forms is a 3-isogeny.

\subsection{2-isogenies}

We can map the Jacobi form to the Legendre form by $\til x = x^{\frac{1}{2}}$, $\til y = y x^{-\frac{1}{2}}$.  It then easily follows that the condition for a curve to have a Jacobi 2-isogenous form is that $D(\beta) = k D(u)$ where $k \in \Z^{+}$.  From the scale dimensions of $u$ and $x$ that can be deduced from (\ref{e5}), it follows that 
the $H_{2}$, $D_{4}$, and $E_{7}$ curves have a 2-isogenous curve of the form
\be
y^2 = x^3 + x\left(\b-\frac{1}{3}\a^2\right) + \a\left(\frac{2}{27}\a^2 - \frac{1}{2}\b\right).
\label{e13}
\ee  
\begin{itemize}
\item The $H_{2}$ isogenous curve is (\ref{e13}) with $\a= M_{2/3}$ and $\b=u+M^2_{2/3}/3$.  It can only have a $\uf(1)$ flavor symmetry.  It would be interesting to identify this sub-maximal mass deformation of the $H_2$ theory by tuning parameters in an asymptotically free Lagrangian N=2 theory.
\item The $D_{4}$ curve with a 2-isogeny has the form (\ref{e13})
with $\alpha = \tau u + M_{2}$, $\beta = u^{2} + M_{4}$, and discriminant
\be
\Delta = (u^{2}+M_{4})^{2}((\tau^{2}-4)u^{2}+2 \tau M_{2}u + (M_{2}^{2}-4M_{4})).
\ee
If we take the special case $M_{4} = (4-\tau^{2})^{-1} M_{2}^{2}$ then we get
the $\spf(1)$ (adjoint hypermultiplet) sub-maximal deformation of the $D_4$ singularity found in \cite{sw9408}.  Presumably demanding the existence of a Seiberg-Witten one-form enforces the $M_4\propto M_2^2$ identification, though this has not been checked.
\item The $E_7$ curve with a 2-isogeny is (\ref{e13})
with $\alpha = M_{2}u + M_{6}$, $\beta = u^{3} + M_{8}u + M_{12}$. The mass parameter dimensions correspond to the dimensions of the Weyl invariants of the exceptional $F_4$ flavor symmetry. A systematic search for the SW 1-form for this curve is in progress \cite{awta}.
\end{itemize}

\subsection{3-isogenies}

A 3-isogenous map from the Hessian form to the Legendre form of the elliptic curve is given by $\til x = -x y^{-\frac{1}{3}}$, $\til y = y^{\frac{1}{3}}$. The resulting condition for a curve to have a Hessian 3-isogenous form is  $D(\gamma) = k D(u)$ where $k \in \Z^{+}$.  The scaling in (\ref{e5}) then shows that the $H_{3}$ and $E_{6}$ curves have a 3-isogenous deformation of the form
\be
y^2 = x^3 - x \delta \left(\gamma +\frac{1}{12}\delta^3\right) + \left(-\frac{1}{108}\delta^6 - \frac{1}{6}\gamma \delta^3 + \gamma^2\right).
\label{e15}
\ee
\begin{itemize}
\item The $H_{3}$ isogenous curve has $\g=u+M_{1/2}^3/12$ and $\d=M_{1/2}$, and can only have a $\uf(1)$ flavor symmetry.  It would be interesting to identify this sub-maximal mass deformation of the $H_3$ theory by tuning parameters in an asymptotically free Lagrangian N=2 theory.
\item The $E_{6}$ curve with a 3-isogeny is (\ref{e15})
with $\delta = M_{2}$, $\gamma = u^{2} + M_{6}$ and discriminant
\be\label{eg2disc}
(u^2 + M_2^3 + M_6)^3 (9 u^2 + 5 M_2^3 + 9 M_6).
\ee  
We have explicitly constructed a SW 1-form for this curve \cite{awta}, following the method of \cite{mn9608}.  We find that the flavor symmetry of this curve is the $G_{2}$ exceptional algebra.
\end{itemize}

\section{Predictions from S-duality}

To summarize so far, we have succeeded in constructing an N=2 supersymmetric sub-maximal mass deformation of the $E_6$ singularity with flavor algebra $\hf=G_2$, have not been able to rule out the possibility of another such deformation with $\hf=\suf(3)$, and have evidence that a sub-maximal deformation of the $E_7$ singularity exists with $\hf=F_4$.  

We would now like to compare these results with the predictions made in \cite{aw0712} for sub-maximal mass deformations of rank-1 SCFTs.  These predictions were based on the fact that N=2 SCFTs can arise as decoupled factors of the strong-coupling limit of certain Lagrangian SCFTs.
In particular, it was argued in \cite{as0711} that 
the physics at infinite coupling of an N=2 Lagrangian SCFT 
with a gauge algebra $\gf$ of rank $r$ is a weakly coupled 
scale-invariant gauge theory with a gauge algebra $\tgf$ with 
smaller rank $s$ which is coupled to an isolated 
rank-$(r{-}s)$ N=2 SCFT.  The coupling between $\tgf$ and the SCFT is the standard gauge coupling: $\tgf$ gauges a subalgebra of the global symmetry
algebra, $\hf$, of the SCFT.  

If one knows the Seiberg-Witten curve and one-form of the $\gf$ Lagrangian SCFT, then one can derive the curve and one-form of the (mass-deformed) isolated SCFT simply by taking the infinite-coupling limit. However, this information (the low energy effective theory on the Coulomb branch) is not know for many theories.\footnote{In light of the solution by D. Gaiotto in \cite{g0904}---and its extension in many following papers---of a very large class of N=2 theories, it may be that sub-maximal mass deformations of some isolated SCFTs may now be able to be derived in this straightforward way.}  In \cite{aw0712} we instead accumulated evidence for the existence of isolated SCFTs and their flavor algebras $\hf$ by assuming a strong-coupling limit of the form described above and testing it with some simple algebraic consistency checks.  In particular, for each potential strong-coupling duality we demanded that on both sides of the duality the spectrum of dimensions of Coulomb branch vevs matched, the flavor symmetry algebras matched, and the number of marginal couplings matched.  We then computed for the isolated SCFT from the presumed duality the flavor algebra central charge $k_{\hf}$, the $\uf(1)_R$ central charge $k_R$, the conformal anomaly $a$, and 
the existence of a global $\Z_{2}$-obstruction to gauging the flavor symmetry.
Finally, we asked whether a given such isolated SCFT was predicted in more than one case.

After examining many examples with one or two marginal couplings, we predicted three new rank 1 sub-maximal mass deformations:
\begin{itemize}
\item The $E_8$ singular curve should have a mass deformation with flavor algebra $\hf=\spf(5)$, a $Z_2$ obstruction, and central charges $k_\hf=7$, $(3/2) k_R=98$, $48a=164$.
\item The $E_7$ singular curve should have a mass deformation with flavor algebra $\hf=\spf(3)\oplus\suf(2)$, a $Z_2$ obstruction in the $\spf(3)$ factor, and central charges $k_{\spf(3)}=5$, $k_{\suf(2)}=8$, $(3/2) k_R=58$, $48a=100$.
\item The $E_6$ singular curve should have a mass deformation with flavor algebra $\hf\neq E_6$ with $r={\rm rank}(\hf)$ between $2\le r\le6$, and central charges $k_{\hf}=(8-n)/{\rm I}$, $(3/2) k_R=38-2n$, $48a=68-2n$, where $n\in\{0,1,2\}$ and ${\rm I}$ is the (positive integer) Dynkin index of embedding of $\suf(2)$ in the maximal embedding $\hf\supset\suf(2)\oplus\uf(1)^p$ with $p\in\{1,2\}$.  If $p=1$ then $n=2$, and if $p=2$ then $n=0$ or $1$.  If $r=2$ then $p=1$, and if $r>2$ then $p=1$ or $2$.
\end{itemize}

The predicted $E_7$ sub-maximal deformation does not match the possible $\hf = F_4$ deformation found as a 2-isogeny.  If would be interesting to see if such a sub-maximal deformation could be located as a factor in the strong-coupling limit of a Lagrangian SCFT by the above methods.

The predicted $E_6$ sub-maximal deformation is loose enough to be compatible with the $\hf=G_2$ deformation constructed as a 3-isogeny.

\section{Central charges and curves}

We can sharpen the above predictions from N=2 S-dualities by a slight refinement \cite{awta} of the arguments of \cite{st0804}.  These arguments show how topologically twisted Seiberg-Witten theory computes the central charges ($k_\hf$, $k_R$, $a$) in terms of the low-energy data ($d$, $h$, $Z$) where
\begin{itemize}
\item $d=D(u)$ is the dimension of the Coulomb branch vev $u$,
\item $h$ is the number of $\uf(1)_{em}$-neutral hypermultiplets at a generic point on the
Coulomb branch ({\it i.e.}, the quaternionic dimension of the Higgs branch above a generic point on the Coulomb branch), and
\item $Z$ is the number of singular points on the Coulomb branch of the curve at generic masses ({\it i.e.}, the number of zeros of the discriminant $\Delta$ counted {\it without} multiplicities).
\end{itemize}
In particular, for rank 1 SCFTs ({\it i.e.}, with one-dimensional Coulomb branches), the relations are $k_{\hf} = 2d - h$, $(3/2) k_R = 2Zd + 4 + 2h$, and 
$48 a = 12d +  2Zd - 2 + 2h$.  The difference between the second two relations gives a known identity, while the new relations can be inverted as
\be\label{e16}
h = 2d-k_\hf, \qquad\mbox{and}\qquad
Z = d^{-1} \left(k_\hf + {3\over4}k_R-2d-2\right),
\ee
to determine the Higgs branch dimension, $h$, and the number of zeros of the discriminant, $Z$.

An $f$-quaternionic-dimensional Higgs branch over generic points on the Coulomb branch on the Lagrangian side of an N=2 S-duality will transform in some representation of the global flavor symmetry $\ff$.  This implies that there must be $h$ hypermultiplets of the dual isolated SCFT factor charged appropriately under its flavor symmetry $\hf$ so that when the relevant $\tgf$ subalgebra of $\hf\supset\tgf\oplus\ff$ is gauged, the $h$ hypermultiplets split into the observed $f$ $\tgf$-gauge-singlet hypermultiplets transforming under $\ff$.  There may be other massless hypermultiplets appearing on the strong-coupling side of the S-duality which are not charged under $\hf$, but they cannot give rise to a Higgs branch over generic points on the Coulomb branch since they enter as otherwise free fields charged under the $\tgf$ gauge group, and so are generically lifted when the $\tgf$ vector multiplet gets a vev.

For example, the $\hf=\spf(5)$ sub-maximal deformation of the $E_8$ singularity appears as a factor in the N=2 S-dual pair \cite{aw0712}:
\be\label{e17}
G_{2} \ {\rm with} \ 8\cdot {\bf 7} \ \simeq \ \suf(2) \ {\rm with} \ ({\bf 2} \oplus {\rm SCFT}[6:\spf(5)]),
\ee
(where SCFT$[6:\spf(5)]$ denotes the $\spf(5)$ sub-maximally deformed CFT).  The values of the central charges imply by (\ref{e16}) that the number of half-hypermultiplets is $2h=10$.  This fills out the pseudoreal ${\bf 10}$ irrep of $\hf=\spf(5)$.  Upon gauging $\tgf=\suf(2)$, $\hf$ is broken as $\spf(5)\supset \suf(2) \oplus \spf(4)$, under which the ${\bf 10}$ decomposes as ${\bf 10}=({\bf 2},{\bf 1}) \oplus ({\bf 1},{\bf 8})$.  This fits perfectly: the $({\bf 2},{\bf 1})$ is the doublet half-hypermultiplet on the right side of (\ref{e17}), while the $ ({\bf 1},{\bf 8})$ are the 8 ``quark" half-hypermultiplets transforming under the $\ff=\spf(4)$ global symmetry apparent on the left side of (\ref{e17}).  These 8 "quark" half-hypermultiplets remain massless at generic points on the Coulomb branch of the $G_2$ theory because the $\bf 7$ representation has a 1-dimensional piece with trivial orbit under the Weyl group of $G_2$.  This also agrees with the predicted $\Z_{2}$-obstruction to gauging the $\hf =\spf(5)$ flavor symmetry of the isolated SCFT: it has a single half-hypermultiplet in a pseudoreal representation.  

Similar remarks apply to the $\hf=\spf(3)\oplus\suf(2)$ sub-maximal deformation of the $E_7$ singularity, which has $2h=6$ half-hypermultiplets transforming in the $\bf 6$ of $\spf(3)$, and none transforming under the $\suf(2)$.

Equation (\ref{e16}) in these two cases also implies that $Z=7$ and $6$, respectively.  This was used in (\ref{e11}) and (\ref{e10}) above to constrain the possible pattern of factorizations of the discriminants of the sub-maximally deformed curves.

Now we use (\ref{e16}) to constrain the possible sub-maximal deformation of the $E_6$ curve mentioned in section 5.  The relevant S-duality in this case is \cite{aw0712}
\be
\suf(3) \ {\rm with} \ {\bf 3}\oplus\bar{\bf 3}\oplus{\bf 6}\oplus\bar{\bf 6} \ \simeq \ \suf(2) \ {\rm with} \ (n\cdot{\bf 2} \oplus {\rm SCFT}[3:\hf]).
\ee
Equation (\ref{e16}) together with the requirements that $h$ and $Z$ be non-negative integers, then gives only 4 solutions:
\be\label{e19}
(n,p,I,Z,h)=
(0,2,2,5,2)\ \mbox{or}\ 
(0,2,8,4,5)\ \mbox{or}\ 
(2,1,1,5,0)\ \mbox{or}\ 
(2,1,2,4,3),
\ee
where $p$ and $I$ are parameters describing the embedding of $\suf(2)\subset\hf$, defined in section 5.  Though this data is still not enough to determine $\hf$, it does put strong constraints on it \cite{awta}.  For example, it rules out the $\hf=G_2$ 3-isogeny deformation, since rank$(G_2)=2$, so $p=1$, and $Z=5$ by (\ref{eg2disc}), picking out the third of the possible solutions in (\ref{e19}). But there is no index $I=1$ $\suf(2)$ subgroup of $G_2$ with commutant $\uf(1)$, so this possibility is ruled out.
It would be interesting to see if the $G_2$ sub-maximally deformed SCFT appeared as a factor in other N=2 S-dualities.


\acknowledgments 
Much of this paper was presented at talks given by the authors at the {\it Great Lakes String} conference at the U. Michigan in March 2009 and at the {\it Quantum Theory and Symmetries 6} conference at U. Kentucky in July 2009.  It is a pleasure to thank D. Gaiotto, A. Shapere, and Y. Tachikawa for helpful discussions.


\begin{thebibliography}{99}
\bibitem{apsw9511}
Argyres P, Plesser R, Seiberg N and Witten E 1996 {\it Nucl. Phys.} B {\bf 461} 71
\bibitem{mn9608} 
Minahan J and Nemeschansky D 1996 {\it Nucl. Phys.} B {\bf 482} 142
\bibitem{sw9408} 
Seiberg N and Witten E 1994 {\it Nucl. Phys.} B {\bf 431} 484
\bibitem{as0711}
Argyres P and Seiberg N 2007 {\it JHEP} {\bf 0712} 088 
\bibitem{aw0712}
Argyres P and Wittig J 2008 {\it JHEP} {\bf 0801} 074
\bibitem{kodaira}
Kodaira K 1963 {\it Ann. Math.} {\bf 77} 563
\bibitem{ad9505}
Argyres P and Douglas M 1995 {\it Nucl. Phys.} B {\bf 448} 93
\bibitem{w82}
Witten E 1982 {\it Phys. Lett.} B {\bf 117} 324
\bibitem{st0804} 
Shapere A and Tachikawa Y 2008 {\it JHEP} {\bf 0809} 109
\bibitem{h04}
see {\it e.g.}, Husem\"oller D 2004 {\it Elliptic Curves}, 2nd ed. (Springer, New York).
\bibitem{awta} Argyres P and Wittig J 2010 to appear
\bibitem{g0904}  
Gaiotto D 2009 N=2 dualities {\it Preprint} arXiv:0904.2715
\end{thebibliography}
\end{document}